\documentclass{aa}             % <-- "class" anstelle von "style"
\usepackage{amssymb,graphicx}  % <-- "graphicx" fuer Bilder, "amssymb"
\begin{document}

\title{The change from accretion via a thin disk to a
coronal flow: dependence on the viscosity of the hot gas}
\author{E. Meyer-Hofmeister and F. Meyer}
\offprints{Emmi Meyer-Hofmeister}
\institute{Max-Planck-Institut f\"ur Astrophysik, Karl-
Schwarzschildstr.~1, D-85740 Garching, Germany
} 

\date{Received:s / Accepted:}

\abstract{
We study the transition from the geometrically thin disk to the hot
coronal flow for accretion onto black holes. The efficiency of
evaporation determines the truncation of the geometrically thin disk
as a function of the black hole mass and the mass flow rate in the
outer disk. The physics of the evaporation was already described
in detail in earlier work (Meyer et al. 2000b). We show now that the value
of the viscosity parameter for the coronal gas has a strong
influence on the evaporation efficiency. For smaller values of the
viscosity evaporation is less efficient. For a given mass flow
rate from outside the geometrically thin disk then extends farther
inward. Spectral transitions between soft and hard states are then
expected for different mass flow rates in the outer disk. The physics is
the same for the cases of stellar and supermassive black holes systems. 
\keywords{accretion disks -- black hole physics  -- galaxies: nuclei
 --  X-rays: galaxies}
}
\titlerunning {The change to a coronal flow: dependence on the viscosity}
\maketitle
%
% 1 figure: H3091F1.ps
v%_____________________________________

\section{Introduction}
For accretion onto black holes at low, sub-Eddington accretion rates
the commonly accepted standard picture is that at larger distance the mass
accretes via a cool geometrically thin, optically thick disk and changes 
at a certain distance to a hot coronal vertically more extended gas
flow. Accretion in this inner region occurs in 
a different mode (advection-dominated accretion flow,
ADAF) where advection rather than radiation removes the locally generated
accretion heat (for reviews see Narayan et al. 1998, Kato et
al. 1998). The physics is the same for accretion onto stellar 
black holes and the black holes in AGN. This picture of 
a cool geometrically thin disk (with a corona above its inner
part), usually modeled using an alpha-viscosity parameterization,
together with the ADAF inside was supported by the spectral fits 
based on this model. In the work of Esin et al. (1997) such a
situation was considered and the spectral transitions observed for
Nova Muscae 1991 were successfully described by a series of 
model disks taking the transition to the ADAF at different distances
from the black hole. Arising from the difficulty to fit also radio and
submillimeter observations it was suggested that the bulk of accreting mass and
energy could be carried off by a wind (Blandford \& Begelmann 1999).
Apart from the question whether wind loss is present or not it is 
important where and in which way the accretion flow changes to the
mode of an advection-dominated flow.

Evaporation of mass from the thin cool disk at the midplane to 
a hot coronal layer above offers a concept to understand the
transition. Such a corona exists
predominantly above the inner region of the cool disk.
The computation of the vertical structure of the corona allows to
determine the amount of mass which flows inward in the corona for each
distance from the black hole or compact star, depending on its mass 
(Meyer \& Meyer-Hofmeister 1994). A detailed description of this
process with application to disks around black holes is given in Meyer
et al. (2000b). With decreasing distance from the black hole more and
more mass flows in the corona. Where the coronal flow reaches 100\% of
the total accretion flow the cool disk in the midplane ends and
inside is only a hot coronal/advection-dominated flow. We consider the
coexistence of thin disk and corona for distances of a few hundred to 
a few thousands of Schwarzschild radii, which applies to the situation
around stellar black holes or nuclei of low-luminosity galaxies. These
are the sources for which the spectra were successfully fitted using
the ADAF model (Narayan et al. 1998).

The evaporation efficiency depends strongly on the viscosity of the
hot coronal gas as already indicated by an analytical approximation formula
(Meyer et al. 2000b). Original computations were carried out for a
standard value $\alpha=0.3$. In the literature various $\alpha$-values
were discussed. The aim of the present paper is
to compute the evaporation efficiency for different viscosity values
in detail and to
evaluate the consequences for the location of the transition.

In Sect. 2 we give a short description of the physics of evaporation
and of our modelling. In Sect. 3 we summarize the viscosity parameters
in the literature. In Sect. 4 we present our results.
In Sect. 5 we summarize our results and discuss which other effects might
also influence the transition.

\section {The physics of evaporation}

The amount of gas which flows in the corona is the amount evaporated
from the cool disk minus wind loss. The equilibrium established between
the hot gas
and the cool disk determines this amount. The coronal evaporation flow
physically results from the following mechanism. The hot corona above
the cool disk atmosphere transports heat downward by electron
conduction. In
the lower layers the temperature decreases from its high coronal value to a
low chromospheric value, heat conduction becomes ineffective, and the
thermal heat flow has to be radiated away. The efficiency of
radiation (in the optically thin case) depends on the square of the
particle number density while the heating is proportional to the
density directly. If this density is too small to radiate
sufficiently, the material will heat up and increase the density in
the corona. In this way a corona of a given temperature (or a given
heating rate) will ``dig'' itself so deep into the chromospheric
layers that a density is reached which is able to radiate away
downward heat conduction, and a stationary state is obtained. This also
establishes the final density in the corona by way of pressure
equilibrium between chromosphere and corona. The gas in the 
corona still has angular momentum and thereby behaves much like an
(albeit thick) accretion disk: Friction transports angular momentum outward and
matter flows inward towards the central object. In the steady state it
is replaced by matter evaporating from the cool disk.

This equilibrium is found computing the vertical structure of the
corona taking into account the relevant processes. We simplify the 
physical two-dimensional situation, for which an exact description is
complicated, additionally difficult due to sonic transition at the
free upper boundary. We describe the evaporation process by
introducing a simplifying one-zone model for the interaction of
corona and disk. This leads to a set of four ordinary differential
equations for the variables temperature, pressure, mass flow and heat
flux (Meyer et al. 2000b) as a model for the equilibrium between cool disk and
corona near the inner edge of the thin disk. Boundary conditions
are (1) at the bottom, $z=z_o$, no heat inflow and a chromospheric
temperature, (2) at the upper boundary no inflow of heat from
infinity and no pressure at infinity (this requires sound transition
at a certain height $z_1$). 
A detailed derivation of the formulae and a description of the
technique of finding the adequate solutions is given in Meyer et al. (2000b).

The evaporation efficiency strongly depends on the mass of the
compact central object. For a given mass the efficiency increases with
decreasing distance to the center (except in an innermost
part). Earlier the coexistence of a cool disk and a corona had already
been applied to disks in dwarf nova systems
(Meyer \& Meyer-Hofmeister 1994, Liu
et al. 1995). The situation is the same in disks around neutron
stars, stellar and supermassive black holes. The quantities accretion
rate and distance can be written in units of Eddington accretion rate
and Schwarzschild radius ($\dot M_{\rm {Edd}}=40{\pi}GM/{\kappa}c$,
$r_{\rm S}=2GM/c^2$, with $M$ mass of central black hole and 
$\kappa$ electron scattering opacity). For the properties of
two-temperature, optically thin ADAFs (Narayan et al. 1998) such an
invariant description was used and the results were successfully applied 
for accretion on stellar black holes and galactic
nuclei. Our relation between evaporation rate and distance also
becomes invariant with the above normalization. The transition to a
coronal flow around massive black holes in galactic nuclei was
discussed in Liu \& Meyer-Hofmeister (2001).

 The compatibility of our modeling of the transition with the
ADAF concept is discussed in Liu et al. (1999). We can expect a 
smooth transition from an thin outer disk + corona to an ADAF.

R\'o\.za\'nska \& Czerny (2000a, 2000b) also investigated the
coexistence of the hot and the cold gas around galactic
black holes and in AGN. Their work follows in
principle the same line as the work of Meyer \& Meyer-Hofmeister
(1994) and Meyer et al. (2000b) on the equilibrium between corona
and thin cool disk around a white dwarf or a black hole, but focuses on the
innermost regions near the black hole.
Though the physical picture is basically the same the results differ
surprisingly in detail: The accretion rate is higher, e.g.
at log $r/r_{\rm{S}}=4$ and for $\alpha=0.1$ their rate is higher by a factor
$\ge$10 than ours. The maximal coronal mass flow rate is reached at
smaller distance from the black hole. R\'o\.za\'nska \& Czerny (2000a,
hereafter RC2000) take a difference between electron and ion
temperature into account and include Compton cooling by soft photons
from the underlying disk. Both effects were neglected in our treatment
because they were unimportant for the coronal structure at distances 
down to that of the maximal evaporation rate. This is confirmed by recent
computations (Liu et al. 2001) including these effects.

Thus the difference between the results must be caused by other 
differences in the approximations. Mainly two facts may be important
in this respect. The semi-analytical approach considers the corona as
one layer of constant pressure and thickness $z=r$. The computed
vertical structure of the corona in Meyer (2000b, Fig.2) has a much
narrower pressure scaleheight, at height $z=r$ the pressure has
dropped already by a factor of 20. This means much less friction and
therefore less heating. A further possible significant difference
appears in the treatment of the energy carried by friction. RC2000 take
the fraction of viscous heating that is caused by advection as
proportional to the ratio of ion temperature to virial temperature
with a factor 1. In our treatment solving the radial diffusion
equation this factor comes out about 5 times larger. The two
differences appear to lead to a significantly cooler corona in our
work and may therefore well account for the difference in the results.
In Fig. 4 of Meyer et al. (2000b) we show a detailed evaluation of the
local energy balance at the distance of maximal evaporation
efficiency. One sees that gains and losses of energy vary
significantly between the bottom of the corona and vertical height $z=r$. It is
therefore difficult to evaluate the changes arising from the approach
of  RC2000 using only one layer of constant pressure and thickness $z=r$.
(Note in this context that there is no difference in the
physical treatment of the lower boundary condition. We include the
full equilibrium between transport of matter and conductive energy 
and radiative losses in our equations. For the lowest region
near the interface between corona and disk chromosphere (energetically
of very little importance) this reduces to the profile that was
already derived by Shmeleva \& Syrovatskii (1973) for the case of the solar
corona.) Considering these arguments it might be preferable to compute
the vertical coronal structure.  

For the innermost region Deufel \& Spruit 
(2001) investigated the
properties of an accretion disk illuminated by ions from the hot
corona.

\section {Values of the hot gas viscosity}
In the literature different values for the $\alpha$-parameter are
discussed. For the ADAF model
fits of the spectra of stellar black holes $\alpha=0.1$, but also
$\alpha=0.3$ were used (Narayan et al. 1996,
Narayan et al. 1997, Esin et al. 1998). For the modeling of the
spectral state transitions of Nova Muscae the value $\alpha$ =0.25
seemed to give the best fit (Esin et al. 1997). Later 
the value  0.1 was used for modeling the spectra of truncated thin disks
in low-luminosity AGN (Quataert et al. 1999). Reasons for the choice
of the $\alpha$ parameter are discussed in Quataert \& Narayan (1999).
Again $\alpha=0.3$ was used for low-radiative-efficiency
accretion in the nuclei of elliptical galaxies (Di Matteo et al. 2000).
R\'o\.za\'nska \& Czerny (2000a) used the value 0.1 for their analysis of the
two-temperature corona. (The parametrization used in
different investigations is not always the same, compare Sect. 5.3).

Another source of knowledge about the viscous stress are
magnetohydrodynamic (MHD) simulations of the inner accretion disk.
It is now becoming increasingly clear that the mechanism of angular
momentum transport is turbulent magnetic stress (Balbus \& Hawley
1998). In their global MHD
simulation Hawley \& Krolik (2001) investigated
the radial structure of disks accreting under influence of MHD turbulent
stresses. From their results they find that $\alpha$ as a function of
radius is below 0.1 in the innermost part and varies between 0.1 and
0.2 further out (their Fig. 13). Note however that this is heavily
weighted for the disk interior and is not characteristic for the
atmosphere where gas pressure is low and the magnetic field is
retively strong (Hawley et al. 2001).
(To really relate
this to the viscosity parameter used for our computations we would
have to consider a magnetic pressure contributing to the ``total''
pressure to a specified degree).

In view of these different values we determined the evaporation
efficiency here for values 0.1 and 0.2 in addition to the earlier results
based on 0.3.

\section {Computational results}

We carried out computations for accretion onto a black hole of
$6 M_\odot$ at distances $r$ from $10^{8.6}$ to $10^{10.5}$ cm,
which corresponds to $10^{2.35}$ to $10^{4.25}$ $r_{\rm S}$,
respectively, with viscosity parameters $\alpha$ 0.1 and 0.2.
In Table 1 we list the new results for a number of distances $r$,
together with the earlier results for the parameter 0.3. $\dot M(r)$ is
the total amount of matter evaporating from the thin disk truncated at
distance $r$.
Mass flow rate $\dot M$ and distance $r$ are also given in units of
Eddington accretion rate and Schwarzschild radius.
The relation between $\dot M$ and $r$
is invariant with respect to changes of the central mass $M$ 
if accretion rates and distances are measured in these
units. The same is true for the coronal temperature (compare Fig. 3
and the comments in Meyer et al. 2000b). The values of
pressure at the lower boundary are listed for 
a discussion of the effect of an outer pressure on the coronal
structure.

\begin{figure}[ht]
\includegraphics[width=8.8cm]{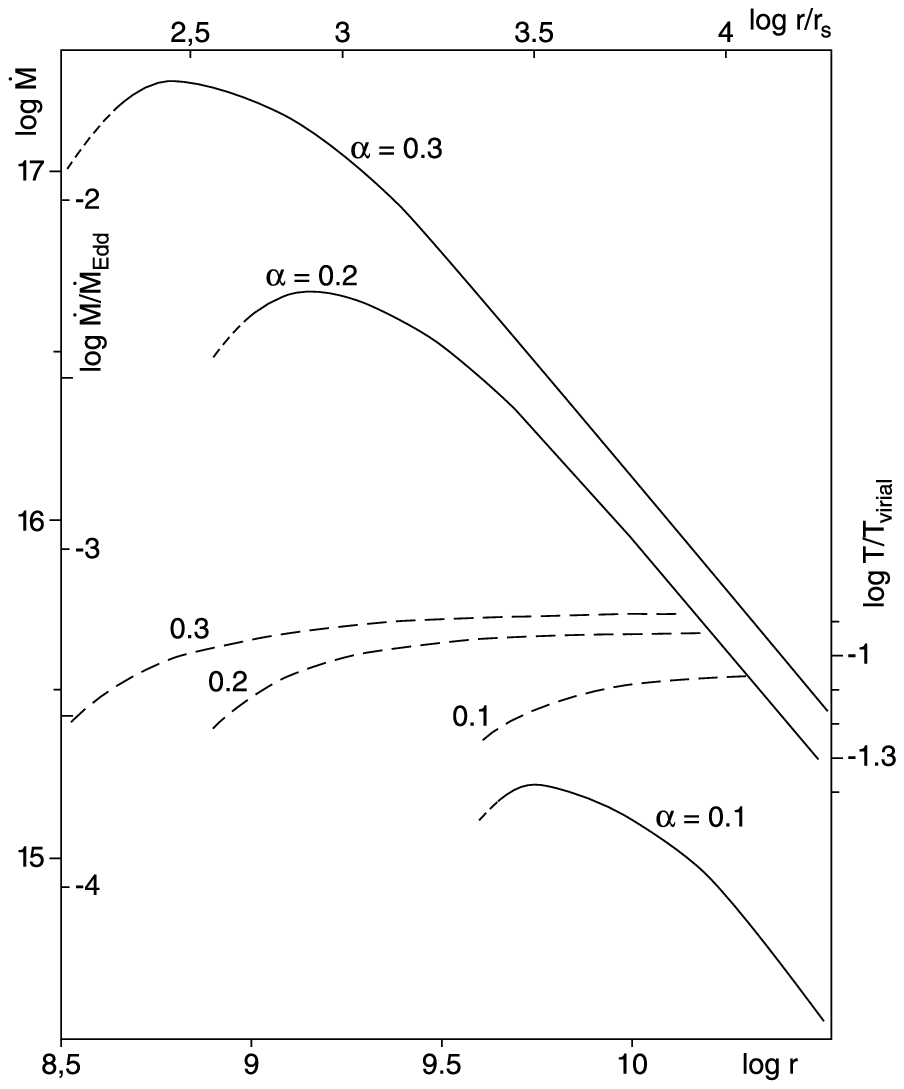}
\caption{Solid lines: Mass evaporation rates  $\dot M$ from a cool disk
with inner edge at distance $r$ from the central object for various
$\alpha$-viscosity values. Rate of inward mass flow in the corona is
$\dot M$ minus wind loss. (Evaporation rate and distances also given in
units of Eddington accretion rate $\dot M_{\rm{Edd}}$ and of
Schwarzschild radius $r_{\rm S}$). The disk is truncated at that
radius where the evaporation rate equals the mass flow rate in the
cool disk. The short-dashed part of
the relations indicates a further decreasing
evaporation efficiency (reliable results there require the consideration of
a two-temperature corona). Dashes lines:
coronal temperatures $T$ at the inner edge of the thin
disk measured in units of virial temperature.
}
\end{figure}

\begin{table*}[ht]
\caption[]{\label{t:evapr-bh} Evaporation efficiency in a disk around 
 a $6M_\odot$ black hole for different values of the viscosity
parameter $\alpha$}
\begin{center}
\begin{tabular}{cccccccc}
\hline
\noalign{\smallskip}
$\alpha$&$\log r$&$\log \dot M=\log(2\pi r^2\dot m_0)$&$\log P_0$&
$\log T_{\rm{max}}$&$\log r/r_{\rm S}$&$\dot M/{\dot M_{\rm Edd}}$\\
\noalign{\smallskip}\hline\noalign{\smallskip} 
0.1&9.6&15.22&5.10&7.91&3.35&$1.99\times10^{-4}$\\
  & 9.8&15.31&4.60&7.83&3.55&$2.45\times10^{-4}$\\
  &10.0&15.22&3.91&7.69&3.75&$1.99\times10^{-4}$\\
  &10.2&14.95&3.16&7.51&3.95&$1.07\times10^{-4}$\\
  &10.5&14.52&1.98&7.24&4.25&$3.97\times10^{-5}$\\
0.2&8.9&16.46&7.75&8.66&2.65&$3.46\times10^{-3}$\\
  & 9.1&16.65&7.31&8.62&2.85&$5.36\times10^{-3}$\\
  & 9.3&16.63&6.70&8.48&3.05&$5.12\times10^{-3}$\\
  & 9.5&16.51&6.03&8.32&3.25&$3.88\times10^{-3}$\\
  & 9.7&16.32&5.30&8.14&3.45&$2.51\times10^{-3}$\\
  &10.0&15.96&4.16&7.85&3.75&$1.09\times10^{-3}$\\
  &10.5&15.29&2.23&7.35&4.25&$2.34\times10^{-4}$\\
0.3&8.5&16.97&9.11&9.04&2.25&$1.13\times10^{-2}$\\
  & 8.8&17.26&8.45&8.97&2.55&$2.16\times10^{-2}$\\
  & 9.0&17.21&7.83&8.82&2.75&$1.95\times10^{-2}$\\
  & 9.3&17.00&6.79&8.57&3.05&$1.20\times10^{-2}$\\
  & 9.7&16.55&5.30&8.19&3.45&$4.28\times10^{-3}$\\
  &10.0&16.15&4.14&7.90&3.75&$1.71\times10^{-3}$\\
  &10.5&15.48&2.21&7.41&4.25&$3.62\times10^{-4}$\\

\noalign{\smallskip}\hline\noalign{\smallskip} 
\end{tabular}
\end{center} 
Notation: $\dot M$ evaporation rate; $\dot m_0$ and $P_0$ vertical mass
flow density and pressure at the lower boundary of the corona,
$T_{\rm{max}}$ temperature at the upper boundary (sound transition); 
quantities $r/r_{\rm S}$ and $\dot M/{\dot M_{\rm Edd}}$ scaled to
Schwarzschild radius and Eddington accretion rate.
$r, \dot m_0, P_0$ in cgs units, $T$ in Kelvin.
\end{table*}

In Fig. 1 we show the
evaporation rate at different distances for the three values of the
viscosity parameter $\alpha$ = 0.1, 0.2, 0.3.
Note that three similar rates are considered: at the inner edge of the
thin disk the evaporation rate is equal to the mass flow rate
(this fact causes the truncation of the thin disk there).
The mass flow rate in the corona is equal to the evaporation rate
minus wind loss. The accretion rate onto the black hole is equal to the
mass flow rate in the corona, if no wind loss  occurs further in. 
The result shows two important changes: (1) The evaporation efficiency
at a given distance $r$ is much lower for a smaller viscosity,
up to a factor of 10 for $\alpha=0.1$ compared to  0.3; (2) for smaller
$\alpha$, the evaporation efficiency
reaches its maximum at larger distances from the black hole, at about
$r=10^{9.2}$cm for $\alpha=0.2$ and $r=10^{9.75}$cm for 0.1
($r=10^{8.8}$cm for $\alpha=0.3$).
This strong dependence on the viscosity parameter is in qualitative
agreement with our earlier analytical estimate (Meyer et al. 2001).
The estimate however was based on a consideration of the global
balance of the various energy terms and also neglected wind losses.

These quite different values for different $\alpha$
indicate that, for a given mass flow rate in the outer thin disk
the disk truncation occurs at very different distances. While, in the
case of a $6 M_\odot$ black hole the mass flow of $10^{-10}M_\odot/yr$
($10^{15.8}$g/s), leads to the disk truncation at 
$10^{10.26}$ cm for $\alpha=0.3$ the thin disk would already extend inward 
all the way towards the last stable orbit for $\alpha=0.1$.
A disk truncation for $\alpha=0.1$ would only be predicted for very low
mass flow rates in the outer disk. a 'hole' in the thin disk would
then occur in much fewer cases. 

Also shown in Fig. 1 is the maximal temperature in the corona
at the inner edge of the thin disk measured in units of the virial
temperature $T_{\rm{virial}}= GM/{(r\cdot\frac{\Re}{\mu}) }$.
We see that from 
distances of $10^{10.3}$cm on outward the ratio becomes almost constant
and approaches values around one tenth. This is a measure for the
pressure scaleheight in the disk, smaller for smaller values $\alpha$.

The fraction of mass carried away by the wind also changes with
the viscosity. For $\alpha=0.3$ the fraction increased
from a very low value at the evaporation efficiency maximum (near log
$r$=8.8) to about 20\% (see fraction $\lambda$ in Table 1, Meyer et
al. 2000b). If we want to compare this with the new results we have to
do this at comparable distances (same factor in radius, relative to the
radius of the evaporation
maximum). The fractions 0.22 at $10^{10.1}$cm (for $\alpha=0.3$) and 0.17 at  
$10^{10.5}$cm ($\alpha=0.2$) show that the part carried away by the wind
decreases with $\alpha$. This corresponds to the lower temperatures
reached and the smaller pressure scaleheight.

\section {Discussion}

\subsection {Further effects on the transition to the hot flow}

The inner accretion regions considered here are in principle
embedded in surrounding gas. If the embedding gas pressure would
be sufficiently high so that the coronal gas pressure can be overcome 
the solution would be significantly affected and both heat and gas
could flow into the corona from above. In Table 1 the values of the pressure at
the transition from disk atmosphere to corona are listed. The pressure
values relevant for the comparison might be a factor of a few smaller
(for an example of coronal structure see Meyer et al. 2000b, Fig.2). 
Cooling flows as discussed in the literature occur at very much larger
distances than discussed here and are hardly relevant in this context.

Of more direct influence however could be a magnetic field that was
carried out with a wind from the innermost ADAF type accretion region
and could spread out like an umbrella over the disk. If its pressure
becomes comparable to the pressure of the coronal evaporation
flow the latter could be suffocated and possibly interesting
alterations on the evaporation transition model may result.

\subsection{Consequences from the new results for the evaporation efficiency}

The question arises whether observations can help to discriminate
between different $\alpha$ values. We shortly summarize here these
studies, first the ones concerning stellar black holes, then
applications to AGN.

For X-ray novae containing a
black hole we determined (using $\alpha=0.3$) the truncation of the
thin disk taking the mass flow rates from the ADAF based spectral fits.
We found the truncation of the thin disk in good agreement with the
location of the inner disk edge inferred from the observed
$H_{\alpha}$ emission line (Meyer et al. 2000b). Also in good agreement
was our prediction for the spectral state transitions (Meyer et
al. 2000a) which happen at X-ray luminosities of about $10^{37}$erg/s
(Tanaka \&Shibazaki 1996) for most sources. A thin disk truncation
based  on the new results for $\alpha=0.1$ would not give a good 
description compared to observation. Also Quataert \& Narayan (1999) had
pointed out that large values, about 0.25, are needed in applications 
of the ADAF model to X-ray binaries. Further support for the choice of
$\alpha=0.3$ comes from the fact that with this evaporation rate the
long outburst cycles of X-ray novae can be understood. The evaporation
is an important feature for the disk evolution during quiescence 
with good agreement between observation and modeling concerning
the amount of matter accumulated in the disk for the next outburst
and the recurrence time as shown in detailed investigations for
A0620-00 (Meyer-Hofmeister \& Meyer 1999, 2001).

Concerning the truncation of the
geometrically thin disks around black holes in galactic nuclei the
situation is not so clear. For M87 and NGC 4649 the truncation
determined from the evaporation efficiency computed
for $\alpha=0.3$ (Liu et al. 1999, Liu \& Meyer-Hofmeister 2001)
gives agreement with
the truncation deduced using the ADAF based spectral fit
(Narayan et al. 1999). But for low-luminosity AGN the observed UV
flux seems to demand a truncation far inward for mass flow rates
roughly similar in both cases (Quataert et al. 1999), which could
not be explained
with a smaller viscosity value in our model: For the discussed
mass flow rate and a small viscosity value 
the disk would instead reach inward towards the last stable orbit and the
spectrum would be soft. This discrepancy might point to further
effects, maybe from a magnetic field, on the transition from
thin disk + corona to the ADAF.

A strong dependence of the evaporation process on the chosen viscosity
parameter was also found by R\'o\.za\'nska \& Czerny (2000a).
They show the change of the evaporation rate for values
$\alpha$ between 0.05 and 0.1. The changes are
qualitatively similar to what we find, but differ from ours as already
discussed in Sect. 2.

\subsection {The viscosity parametrization}

Discussing the numerical values for the viscosity parameter one should keep
in mind the following aspects.

(1) The definition of the viscosity parametrization differs in different
investigations. In the original formulation of Shakura \& Sunyaev (1973)
the viscous stress is given as $\alpha {V_{\rm s}}^2  \rho$
($ V_{\rm s}$ sound velocity, $\rho$ density). For Keplerian rotation
in the disk the viscous stress is $\frac{3}{2} \mu \Omega_{\rm K}$ which
leads to $\mu$=$\frac{2}{3} \alpha  {V_{\rm s}}^2 \rho  \Omega_{\rm K}$ for
the dynamical and to $\nu$=$\frac{2}{3} \alpha 
{V_{\rm s}}^2  \Omega_{\rm K}$ for the kinematic viscosity.
Frank et al. (1992) choose  $\nu$=$\alpha {V_{\rm s}}^2 
\Omega_{\rm K}$, the ansatz without the factor $\frac{2}{3}$.
The latter formula
is used for the computations of the ADAF in the work of Narayan
et al. (1997), which means e.g. a value $\alpha$=0.2 there would correspond
to a value 0.3 in our investigations. To relate the results of the MHD
simulations to $\alpha$ parameter values the Shakura-Sunyaev
formulation was used in Hawley \& Krolik (2001).

(2) With the acceptance of the magnetic nature of
friction in accretion disks the $\alpha$ value should be related to
the magnetic pressure. The value $\beta$, the ratio of gas
pressure to magnetic pressure should account for this. The results for
otherwise the same parameter $\alpha$ are comparable only if also this
value is the same. We have not included the magnetic pressure
explicitely in our analysis. 
Our present calculation confirms the strong influence that the value
of $\alpha$ has on the maximal evaporation rate and the radial
position of this maximum. The near quantitative agreement with the
analytical estimate of this effect by Meyer et al. (2000b, equ. (57)
which neglects the effects of wind loss) lends some credibility also
to the estimate of the effect that an inclusion of magnetic fields in
the way suggested in that paper would have. Hawley et al. (2001) have
shown that hydrodynamic simulations indicate a value of $\beta>1$ in
the main body of the accretion disk that formed, but $\beta<1$ in the
atmosphere where gas pressure drops off on a much shorter scale than
magnetic pressure. Taking $\beta$=0.5 as an example and reducing
the effective thermal conductivity $\kappa_0$ by a mean value
$\cos^2\vartheta$=0.25 for the field inclination $\vartheta$
as discussed there this
would compensate nearly exactly the effect on the maximal evaporation
rate of decreasing $\alpha$ by a factor 3 and at the same time shift
the radius of maximal evaporation even further in. Due to the strong
dependencies one must be very careful when comparing different
numerical values for $\alpha$ given in the literature.

\begin{acknowledgements}

\end{acknowledgements}

\end{document}